\documentclass[12pt,reqno]{amsart}

\usepackage{bm}
\usepackage{amssymb}

\newtheorem{theorem}{Theorem}

\newtheorem{lemma}{Lemma}

\newcommand{\conn}{\overline {\nabla}}

\newcommand{\wD}{\widetilde{D}}
\newcommand{\oD}{\overline{D}}

\newcommand{\s}{\mathbb{S}}

\newcommand{\T}{\mathcal{T}}

\newfont{\bb}{msbm10 at 12pt}

\def\pf{{\textit {Proof :} }}
\def\qed{\hfill{Q.E.D.}\smallskip}

\newcommand{\ls}{\setlength{\baselineskip}{12pt}
                 \setlength{\parskip}{3mm}}

\newcommand{\mysection}[1]{\section{#1}\setcounter{equation}{0}}

\newcommand{\bal}{\begin{align}}      \newcommand{\eal}{\end{align}}
\newcommand{\ba}{\begin{array}}      \newcommand{\ea}{\end{array}}
\newcommand{\bc}{\begin{center}}     \newcommand{\ec}{\end{center}}
\newcommand{\be}{\begin{enumerate}}  \newcommand{\ee}{\end{enumerate}}
\newcommand{\beq}{\begin{eqnarray}}  \newcommand{\eeq}{\end{eqnarray}}
\newcommand{\beQ}{\begin{eqnarray*}} \newcommand{\eeQ}{\end{eqnarray*}}
\newcommand{\bi}{\begin{itemize}}    \newcommand{\ei}{\end{itemize}}
\newcommand{\bt}{\begin{tabular}}    \newcommand{\et}{\end{tabular}}
\newcommand{\bdm}{\begin{displaymath}} \newcommand{\edm}{\end{displaymath}}

\begin{document}

\title[quasi-local mass]{A quasi-local mass for 2-spheres with negative Gauss curvature }

\address{Institute of Mathematics,
Academy of Mathematics and Systems Science, Chinese Academy of
Sciences, Beijing 100080, China}

\author{Xiao Zhang}
\thanks{Partially supported by NSF of China(10421001), NKBRPC(2006CB805905) and the
Innovation Project of Chinese
Academy of Sciences.}

\email{xzhang@amss.ac.cn}

\begin{abstract}
We extend our previous definition of quasi-local mass to 2-spheres
whose Gauss curvature is negative and prove its positivity.
\end{abstract}

\keywords{General relativity, quasi-local mass, positivity}

\subjclass[2000]{53C27, 53C50, 83C60}

\maketitle

\mysection{Introduction} \ls

In \cite{LY1}, Liu and Yau propose a definition of quasi-local mass
for any smooth spacelike, topological 2-sphere with positive Gauss
curvature. In particular, Liu and Yau \cite{LY1, LY2} are able to
use Shi-Tam's result \cite{ST1} to prove its positivity. When the
Gauss curvature of a 2-sphere is allowed to be negative, Wang and
Yau \cite{WY} use Pogorelov's result \cite{P} to embed the 2-sphere
into the hyperbolic space to generalize Liu-Yau's definition, and
prove its positivity by using a spinor argument of the positive mass
theorem for asymptotically hyperbolic manifolds \cite{W, CH, Z1}.
Wang-Yau's result is improved in certain sense by Shi and Tam
\cite{ST2}.

In attempting to resolve the decreasing monotonicity of
Brown-York's quasi-local mass \cite{BY1, BY2}, the author
\cite{Z2} propose a new quasi-local mass and prove its positivity
essentially for 2-spheres with positive Gauss curvature. It is
still open when the 2-spheres have nonnegative Gauss curvature
because the isometric embedding into $\mathbb{R} ^3$ in this case
is only proved to be $C^{1,1}$ by Guan-Li and Hong-Zuily \cite{GL,
HZ}. However, we expect the $C ^{1,1}$ regularity is sufficient
for our propose, and we address it elsewhere.

In this note, we use the idea of Wang and Yau to extend the
quasi-local mass in \cite{Z2} to the case of 2-spheres with
negative Gauss curvature. We embed such 2-spheres into the
(spacelike) hyperbola in the Minkowski spacetime which has the
nontrivial second fundamental form. By using the constant spinors
in the Minkowski spacetime, we can solve a boundary problem for
the Dirac-Witten equation. Then, the method in \cite{Z2} gives
rise to the quasi-local mass as well as its positivity. We would
like to point out that our quasi-local mass is only one quantity,
while the one defined by Wang and Yau is a 4-vectors. This
difference is due to the hyperbola in our approach goes to null
infinity in the Minkowski spacetime, and the one in Wang-Yau's
approach goes to spatial infinity in the Anti-de Sitter spacetime,
which has trivial second fundamental form. The positive mass
theorem near null infinity in asymptotically Minkowski spacetimes
was established in \cite{Z1, Z1-1}.

\mysection{Dirac-Witten equations} \ls

In this section, we will review the existences of the Dirac-Witten
equations proved in \cite{Z2}. Let $(N, {\widetilde g})$ be a
4-dimensional spacetime which satisfies the Einstein fields
equations. Let $(M, g, p)$ be a smooth {\it initial data set}. Fix
a point $p \in M$ and an orthonormal basis $\{ e _{\alpha}\}$ of
$T_p N$ with $e_0$ future-time-directed normal to $M$ and $e_i$
tangent to $M$ ($1\leq i \leq 3$).

Denote by $\s $ the (local) spinor bundle of $N$. It exists globally
over $M$ and is called the hypersurface spinor bundle of $M$. Let
$\widetilde \nabla$ and $\conn$ be the Levi-Civita connections of
$\widetilde g$ and $g$ respectively,  the same symbols are used to
denote their lifts to the hypersurface spinor bundle. There exists a
Hermitian inner product $(\;,\;)$ on $\s$ along $M$ which is
compatible with the spin connection $\widetilde \nabla $. The
Clifford multiplication of any vector $\widetilde X$ of $N$ is
symmetric with respect to this inner product. However, this inner
product is not positive definite and there exists a positive
definite Hermitian inner product defined by $\langle\; ,\;\rangle =
(e _0 \cdot\; ,\;)$ on $\s$ along $M$.

Define the second fundamental form of the initial data set $p
_{ij}= \widetilde g (\widetilde \nabla _i e_0, e_j)$. Suppose that
$M$ has boundary $\Sigma $ which has finitely many connected
components $\Sigma ^1, \cdots, \Sigma ^l$, each of which is a
topological 2-sphere, endowed with its induced Riemannian and spin
structures. Fix a point $p \in \Sigma$ and an orthonormal basis
$\{ e _i\}$ of $T _p M$ with $e _r =e _1$ outward normal to
$\Sigma $ and $e _a$ tangent to $\Sigma$ for $2 \leq a \leq 3$.
Let $h _{ab} = \langle \conn _a e _r, e _b \rangle$ be the second
fundamental form of $\Sigma $. Let $H = tr( h )$ be its mean
curvature. $\Sigma $ is a {\it future/past apparent horizon} if
 \begin{eqnarray}
H \mp tr (p | _{\Sigma }) \geq 0 \label{horizon}
 \end{eqnarray}
holds on $\Sigma $. When $\Sigma $ has multi-components, we require
that (\ref{horizon}) holds (with the same sign) on each $\Sigma _i$.
The spin connection has the following relation
 \begin{eqnarray}
\widetilde \nabla _a =\nabla _a +\frac{1}{2}h _{ab} e _r \cdot e _b
\cdot -\frac{1}{2}p _{aj} e _0 \cdot e _j \cdot. \label{two-conn3}
 \end{eqnarray}

The Dirac-Witten operator along $M$ is defined by $\widetilde D =
 e _i \cdot \widetilde \nabla _i $. The Dirac operator of $M$ but
acting on $\s$ is defined by $\oD = e _i \cdot \conn _i $. Denote by
$\nabla $ the lift of the Levi-Civita connection of $\Sigma $ to the
spinor bundle $\s | _{\Sigma}$. Let $D=e _a \cdot \nabla _a $ be the
Dirac operator of $\Sigma$ but acting on $\s | _{\Sigma}$. The
Weitzenb{\"o}ck type formula gives rise to
 \begin{eqnarray}
& &\int _M  |\widetilde \nabla \phi |^2 + \langle \phi, \T \phi
\rangle - |\widetilde D \phi |^2  \nonumber\\
 &=&\int _{\Sigma} \langle \phi, (e _r\cdot D -\frac{H}{2}+\frac{tr (p | _{\Sigma
 })}{2} e _0\cdot e _ r\cdot
-\frac{p _{ar}}{2} e _0 \cdot e _a \cdot )\phi\rangle . \label{w2}
 \end{eqnarray}
where $\T =\frac{1}{2}(T _{00} + T _{0i} e  _0 \cdot e _i\cdot)$. If
the spacetime satisfies the {\it dominant energy condition}, then
$\T$ is a nonnegative operator. Let
 \beQ
P _{\pm}=\frac{1}{2}(Id \pm e _0 \cdot e _r \cdot )
 \eeQ
be the projective operators on $\s | _{\Sigma}$. In \cite{Z2}, we
prove the following existences:
 \bi
\item[(i)] If $tr _g (p) \geq 0$ and $\Sigma$ is a past apparent
horizon, then the following Dirac-Witten equation has a unique
smooth solution $\phi \in \Gamma (\s)$
 \begin{eqnarray}
 \left\{ \begin{array}{ccccc}
       \wD \phi &=&  0 &in & M\\
  P _{+} \phi &=& P _{+}  \phi _0 & on &\Sigma _{i _0} \\
  P _{+} \phi &=& 0 & on &\Sigma _{i}\;(i \neq i _0)\\
    \end{array} \right . \label{existence1}
 \end{eqnarray}
for any given $\phi _0 \in \Gamma (\s \big| _{\Sigma})$ and for
fixed $i _0 $;
\item[(ii)] If $tr _g (p) \leq 0$ and $\Sigma$ is a future apparent
horizon, then the following Dirac-Witten equation has a unique
smooth solution $\phi \in \Gamma (\s)$
 \begin{eqnarray}
 \left\{ \begin{array}{ccccc}
       \wD \phi &=&  0 &in & M\\
  P _{-} \phi &=& P _{-}  \phi _0 & on &\Sigma _{i _0} \\
  P _{-} \phi &=& 0 & on &\Sigma _{i}\;(i \neq i _0)\\
    \end{array} \right . \label{existence2}
 \end{eqnarray}
for any given $\phi _0 \in \Gamma (\s \big| _{\Sigma})$ and for
fixed $i _0 $.
 \ei

\mysection{Embedding 2-spheres} \ls

Let $(M, g, p)$ be a smooth {\it initial data set} where $M$ has
boundary $\Sigma $ which has finitely many connected components
$\Sigma _1, \cdots, \Sigma _l$, each of which is a topological
2-sphere. Suppose that some $\Sigma _{i_0} $ can be smoothly
isometrically embedded into a smooth spacelike hypersurface
$\breve{M} ^3 $ in the Minkowski spacetime $\mathbb{R} ^{3,1}$ and
denote by $\aleph$ the isometric embedding. Let $\breve{\Sigma}
_{i_0}$ be the image of $\Sigma _{i_0}$ under the map $\aleph$.
Let $\breve{e}_r$ the unit vector outward normal to
$\breve{\Sigma} _{i_0}$ and $\breve{h}_{ij}$, $\breve{H}$ are the
second fundamental form, the mean curvature of $\breve{\Sigma}
_{i_0}$ respectively. Denote by $p_0 = \breve{p} \circ \aleph$, $H
_0 = \breve{H} \circ \aleph $ the pullbacks to $\Sigma$.

The isometric embedding $\aleph$ also induces an isometry between
the (intrinsic) spinor bundles of $\Sigma _{i _0}$ and
$\breve{\Sigma} _{i _0}$ together with their Dirac operators which
are isomorphic to $e _r \cdot D$ and $\breve{e} _r \cdot \breve{D}$
respectively. This isometry can be extended to an isometry over the
complex 2-dimensional sub-bundles of their hypersurface spinor
bundles. Denote by ${\breve{\s} ^{\breve{\Sigma} _{i _0}}}$ this
sub-bundle of $\breve{\s} | _{\breve{\Sigma} _{i _0}}$. Let
$\breve{\phi}$ be a constant section of ${\breve{\s}
^{\breve{\Sigma} _{i _0}}}$ and denote $\phi _0 = \breve{\phi} \circ
\aleph $. Denote by $\breve{\Xi}$ the set of all these constant
spinors $\breve{\phi}$ with the unit norm. This set is isometric to
$S ^3$.

Let $\breve{D}$ be the (induced) Dirac operator on $\breve{\Sigma}
_{i _0}$ which acts on the hypersurface spinor bundle $\breve{\s}$
of $\breve{M}$. Let $\breve{\phi}$ be the covariant constant spinor
of the trivial spinor bundle on $\mathbb{R} ^{3, 1}$ with unit norm
taking by the positive Hermitian metric on $\breve{\s}$. Then
(\ref{two-conn3}) implies
 \beQ
\breve{\nabla} _a \breve{\phi} +\frac{1}{2}\;\breve{h}
 _{ab} \breve{e} _r \cdot \breve{e} _b \cdot
\breve{\phi} -\frac{1}{2}\; \breve{p} _{aj} \breve{e} _0 \cdot
\breve{e} _j \cdot \breve{\phi} =0
 \eeQ
over $\breve{\Sigma} _{i _0}$. Pullback to $\Sigma _{i _0}$, we
obtain
 \beq
e _r \cdot D \phi _0=\frac{H_0}{2} \phi _0 -\frac{1}{2} p _{0aa} e
_0 \cdot e _r \cdot \phi _0 +\frac{1}{2}p _{0ar} e _0 \cdot e _a
\cdot \phi _0   \label{Hp}
 \eeq
over $\Sigma _{i _0}$. Denote $\phi _0 ^\pm = P _{\pm} \phi _0$.
Since $ e _r \cdot D \circ P _{\pm} =P _{\mp} \circ e _r \cdot D, $
(\ref{Hp}) gives rise to
 \beQ
e _r \cdot D \phi ^+ _0 &=& \frac{H_0}{2} \phi ^- _0 +\frac{1}{2}p
_{0aa} \phi ^- _0
+\frac{1}{2}p _{0ar} e _0\cdot e _a \cdot \phi ^+ _0,\\
e _r \cdot D \phi ^- _0 &=& \frac{H_0}{2}\phi ^+ _0 -\frac{1}{2}p
_{0aa} \phi ^+ _0 +\frac{1}{2}p _{0ar} e _0\cdot e _a \cdot \phi ^-
_0.
 \eeQ
Therefore, using
 \beQ
\int _{\Sigma _{i _0}} \langle \phi ^- _0, e _r \cdot D \phi ^+ _0
\rangle =\int _{\Sigma _{i _0}} \langle e _r \cdot D\phi ^- _0, \phi
^+ _0\rangle,
 \eeQ
we obtain
  \beq
\int _{\Sigma _{i _0}} (H _0 -p _{0aa} )|\phi _0 ^+ | ^2 =\int
_{\Sigma _{i _0}}(H _0 +p _{0aa} ) |\phi _0 ^- | ^2. \label{w-phi0}
 \eeq

In this paper, we introduce the following conditions on $M$:
 \bi
 \item[(i)] $tr _g(p) \geq 0$, $H | _{\Sigma _i} + tr (p | _{\Sigma
_i}) \geq 0$ for all $i$;
 \item[(ii)] $tr _g(p) \leq 0$, $H |
_{\Sigma _i} - tr (p | _{\Sigma _i}) \geq 0$ for all $i$.
 \ei

\begin{lemma} \label{lemma}
Let $(N ^{3,1}, {\widetilde g})$ be a spacetime which satisfies
the dominant energy condition. Let $(M, g, p)$ be a smooth
spacelike (orientable) hypersurface which has boundary $\Sigma$
with finitely many multi-components $\Sigma _i$, each of which is
a topological sphere. Suppose that $\Sigma _{i _0}$ can be
smoothly isometrically embedded into some spacelike hypersurface
$(\breve{M}, \breve{g}, \breve{p})$ in the Minkowski spacetime
$\mathbb{R} ^{3,1}$. Let $\aleph$ be the isometric embedding and
let $\breve{\Sigma} _{i_0}$ be the image of $\Sigma _{i_0}$.
Suppose either condition $(i)$ holds and $\breve{\Sigma} _{i_0}$
are past apparent horizons, i.e.,
$$\breve{H}+tr(\breve{p} | _{\breve{\Sigma}_{i _0}}) \geq 0,$$ or
condition $(ii)$ holds and $\breve{\Sigma} _{i_0}$ are future
apparent horizons, i.e.,
$$\breve{H}-tr(\breve{p} | _{\breve{\Sigma}_{i _0}}) \geq 0.$$ Let
$\phi $ be the unique solution of (\ref{existence1}) or
(\ref{existence2}) for some $\breve{\phi} \in \breve{\Xi}$. Then
 \beQ
\int _{\Sigma _{i_0}} \langle \phi,e _r \cdot D \phi\rangle \leq
\frac{1}{2}\int _{\Sigma _{i_0}} \langle \phi, (H _0 -p _{0aa}e_0
\cdot e _r \cdot +p _{0ar} e_0 \cdot e _a \cdot )\phi \rangle.
 \eeQ
\end{lemma}
\pf Assume condition $(i)$ holds and $\breve{\Sigma} _{i_0}$ are
past apparent horizons. Let $\phi $ be the smooth solution of
(\ref{existence1}) with the prescribed $\phi _0$ on $\Sigma _{i_0}$.
Denote $\phi ^\pm=P _{\pm} \phi$. Denote $\phi _0 ^\pm =P _{\pm}
\phi _0$. By the boundary condition, we have $\phi ^+ =\phi _0 ^+$.
Thus
 \beQ
\int _{\Sigma _{i_0}} \langle \phi,e _r \cdot D \phi \rangle &=& 2
\Re \int _{\Sigma _{i_0}} \langle \phi ^-,e _r \cdot D \phi _0
^+\rangle \\
&=& \Re \int _{\Sigma _{i_0}} \langle \phi ^-, H _0 \phi _0 ^- + p
_{0aa} \phi _0 ^- +p _{0ar} e _0 \cdot e _r \cdot \phi _0 ^+ \rangle \\
&\leq &\frac{1}{2}\int _{\Sigma _{i_0}} (H _0 +p _{0aa})(|\phi
^- | ^2+|\phi _0 ^- | ^2 )\\
& & + \Re \int _{\Sigma _{i_0}} \langle \phi ^-,p _{0ar} e _0\cdot e
_a \cdot \phi ^+ _0 \rangle \\
&=& \frac{1}{2} \int _{\Sigma _{i_0}} (H _0 +p _{0aa} ) |\phi
^- | ^2 + (H _0 -p _{0aa} )|\phi _0 ^+ | ^2 \\
& &+\Re \int _{\Sigma _{i_0}} \langle \phi ^-,p _{0ar} e _0\cdot e
_a
\cdot \phi ^+ \rangle \\
&=&\frac{1}{2} \int _{\Sigma _{i_0}} H _0 |\phi | ^2 +p _{0aa}
(|\phi ^-| ^2 -|\phi ^+ | ^2 ) \\
& &+\Re \int _{\Sigma _{i_0}} \langle \phi ^-,p _{0ar} e _0\cdot e
_a
\cdot \phi ^+ \rangle. \\
 \eeQ
Note that
 \beQ
\langle \phi,p _{0aa} e _0\cdot e _r \cdot \phi \rangle = p _{0aa}
(|\phi ^+ | ^2 -|\phi ^- | ^2  ).
 \eeQ
Moreover, that $e _0 \cdot e _a \cdot P _{\pm} = P _{\mp} \cdot e _0
\cdot e _a \cdot $ gives rise to
 \beQ
\langle \phi,p _{0ar} e _0\cdot e _a \cdot \phi \rangle =2 \Re
\langle \phi ^-,p _{0ar} e _0\cdot e _a \cdot \phi ^+ \rangle.
 \eeQ
Same argument is applied under condition $(ii)$. We finally prove
the lemma. \qed

\mysection{Quasi-local mass} \ls

Now we use the idea of Wang and Yau \cite{WY} (see also \cite{ST2})
to extend the definition of quasi-local mass in \cite{Z2} to the
case of 2-spheres with negative Gauss curvature.

We first review the definition for 2-spheres with nonnegative Gauss
curvature in \cite{Z2}: Suppose some $\Sigma _{i_0}$ can be smoothly
isometrically embedded into $\mathbb{R} ^3 $ in the Minkowski
spacetime $\mathbb{R} ^{3,1}$ and denote $\breve{\Sigma} _{i_0}$ its
image. (It exists if $\Sigma _{i_0} $ has positive Gauss curvature.)
In this case, $\breve{p}=0$.

Let $\phi $ be the unique solution of (\ref{existence1}) or
(\ref{existence2}) for some $\breve{\phi} \in \breve{\Xi}$. Denote
 \beq
m(\Sigma _{i _0}, \breve{\phi})&=&\frac{1}{8\pi}\Re \int _{\Sigma
_{i_0}}
\Big[(H _0 -H ) | \phi | ^2 \nonumber\\
& & + tr(p |_{\Sigma _{i_0}}) \langle \phi, e_0 \cdot e _r \cdot
\phi \rangle  \nonumber\\
& &- p _{ar} \langle \phi, e_0 \cdot e _a \cdot \phi \rangle \Big].
\label{pre-local-mass}
 \eeq
The {\it quasi local mass of $\Sigma _{i _0}$} is defined as
 \beq
m(\Sigma _{i _0})=\min _{\breve{\Xi}}m(\Sigma _{i _0},
\breve{\phi}). \label{local-mass}
 \eeq
If all $\Sigma _i $ can be isometrically embedded into $\mathbb{R}
^3 $ in the Minkowski spacetime $\mathbb{R} ^{3,1}$, we define the
{\it quasi local mass of $\Sigma $} as
 \beq
m(\Sigma )=\sum _i m(\Sigma _{i}). \label{total-local-mass}
 \eeq

If the mean curvature of $\breve{\Sigma} _{i_0}$ is further
nonnegative (it is true if $\Sigma _{i_0} $ has positive Gauss
curvature), we can prove the positivity of the quasi-local mass
(\ref{local-mass}) (Theorem 1 in \cite{Z2}).

Now suppose some $\Sigma _{i_0}$ has negative Gauss curvature and
let $$K _{\Sigma _{i_0}} \geq -\kappa ^2$$ ($\kappa > 0$) where
$-\kappa ^2$ is the minimum of the Gauss curvature. (Here we must
choose the minimum of the Gauss curvature instead of arbitrary lower
bound, otherwise the quasi-local mass defined in the following way
might depend on this arbitrary lower bound.) By \cite{P, CW},
$\Sigma _{i_0}$ can be smoothly isometrically embedded into the
hyperbolic space $\mathbb{H} _{-\kappa ^2} ^3 $ with constant
curvature $-\kappa ^2$ as a convex surface which bounds a convex
domain in $\mathbb{H} _{-\kappa ^2} ^3 $. Let $(t, x_1, x_2, x_3)$
be the spacetime coordinates of $\mathbb{R} ^{3,1}$. Then
$\mathbb{H} _{-\kappa ^2} ^3 $ is one-fold of the spacelike
hypersurfaces
 \beQ
\big\{(t, x_1, x_2, x_3) \big | t^2 -x _1 ^2 -x_2 ^2 -x _3 ^2 =
\frac{1}{\kappa ^2} \big\}.
 \eeQ
The induced metric of $\mathbb{H} _{-\kappa ^2} ^3 $ is
 \beQ
\breve{g} _{\mathbb{H} _{-\kappa ^2} ^3 } =\frac{1}{1+ \kappa ^2 r
^2} dr ^2+r ^2 (d \theta ^2 +\sin ^2 \theta d \psi ^2 )
 \eeQ
It has the second fundamental form $\breve{p} ^{+} _{\mathbb{H}
_{-\kappa ^2} ^3 }= \kappa \breve{g} _{\mathbb{H} _{-\kappa ^2} ^3
}$ for the upper-fold $\{t
>0\}$ and $\breve{p} ^{-} _{\mathbb{H} _{-\kappa ^2} ^3 }= -\kappa \breve{g}
_{\mathbb{H} _{-\kappa ^2} ^3 }$ for the lower-fold $\{t <0\}$ with
respect to the future-time-directed normal. Denote also
$\breve{\Sigma} _{i_0}$ its image.

Let $\phi $ be the unique solution of (\ref{existence1}) or
(\ref{existence2}) for some $\breve{\phi} \in \breve{\Xi}$. Denote
 \beq
{\hat m} _{\pm}(\Sigma _{i _0}, \breve{\phi})&=&\frac{1}{8\pi}\Re
\int _{\Sigma _{i_0}}
\Big[(H _0 -H ) | \phi | ^2 \nonumber\\
& & -\big( tr(p _0 |_{\Sigma _{i_0}})- tr(p |_{\Sigma
_{i_0}})\big)\langle \phi, e_0 \cdot e _r \cdot
\phi \rangle  \nonumber\\
& & +(p _{0 ar}- p _{ar})\langle \phi, e_0 \cdot e _a \cdot \phi
\rangle \Big] \label{pre-local-mass-new}
 \eeq
where
\[
 p _{0} =\left\{\begin{array}
         {r@{\;\;}l}
 \mbox{pullback of $\breve{p} ^{+} _{\mathbb{H} _{-\kappa ^2} ^3 }$}: & \mbox{if $\Sigma _{i_0}$ is
 isometrically embedded into}  \\ & \mbox{the upper-fold $\{t>0\}$},\\
 \mbox{pullback of $\breve{p} ^{-} _{\mathbb{H} _{-\kappa ^2} ^3 }$}: & \mbox{if $\Sigma _{i_0}$ is
 isometrically embedded into} \\ & \mbox{the lower-fold $\{t<0\}$}.
  \end{array} \right.
\]
It is easy to see that $tr(p _0 |_{\Sigma _{i_0}})=\pm 2$, thus
 \beQ
{\hat m} _{\pm}(\Sigma _{i _0}, \breve{\phi}) &=&\frac{1}{8\pi}\Re
\int _{\Sigma _{i_0}} \Big[(H _0 -H ) | \phi | ^2 \\
& &+ tr(p |_{\Sigma _{i_0}})\langle \phi, e_0 \cdot e _r \cdot \phi
\rangle \\
& &- p _{ar}\langle \phi, e_0 \cdot e _a \cdot \phi \rangle \Big]\\
& &\mp \frac{\kappa}{4\pi}\int _{\Sigma _{i_0}} \langle \phi, e_0
\cdot e _r \cdot \phi \rangle.
 \eeQ

Now we define the quasi local mass of $\Sigma _{i _0}$ under
conditions $(i)$, $(ii)$ which are introduced in the previous
section.

If condition $(i)$ holds, we embed $\Sigma _{i _0}$ into upper-fold
$\{t>0\}$. Since $\breve{\Sigma} _{i_0}$ is convex, we have
$$\breve{H}+tr(\breve{p} | _{\breve{\Sigma}_{i _0}}) > 0.$$
If condition $(ii)$ holds, we embed $\Sigma _{i _0}$ into lower-fold
$\{t<0\}$. We have
$$\breve{H}-tr(\breve{p} | _{\breve{\Sigma}_{i _0}})> 0$$ in this case.

The {\it quasi local mass} of $\Sigma _{i _0}$ is defined as
 \beq
{\hat m}(\Sigma _{i _0})=\left\{\begin{array}
                  {r@{:\;\;}l}
\min _{\breve{\Xi}}{\hat m} _{+}(\Sigma _{i _0}, \breve{\phi}) &
\mbox{if
condition $(i)$ holds}, \\
\min _{\breve{\Xi}}{\hat m} _{-}(\Sigma _{i _0}, \breve{\phi}) &
\mbox{if condition $(ii)$ holds}.
 \end{array} \right.
\label{local-mass-new}
 \eeq

Note that it might have two different values via embedding to the
upper-fold and to the lower-fold respectively when $tr (p) =0$.
However, since $\wD \phi =0$, $\wD (e _0 \cdot \phi ) = -tr _g (p)
\phi =0$, we have
 \beQ
\int _{\Sigma } \langle e _r \cdot \phi , e _0 \cdot \phi \rangle
= \int _M \langle \wD \phi, e _0 \cdot \phi \rangle - \langle
\phi, \wD (e _0 \cdot \phi)\rangle =0.
 \eeQ
This implies ${\hat m} _{+}(\Sigma _{i _0}, \breve{\phi})={\hat m}
_{-}(\Sigma _{i _0}, \breve{\phi})$. Hence ${\hat m}(\Sigma _{i
_0})$ is unique in this case. Furthurmore, (\ref{local-mass-new})
approaches (\ref{local-mass}) when $\kappa \rightarrow 0$.

If $\Sigma _1, \cdots, \Sigma _{l _0}$ can be isometrically
embedded into $\mathbb{R} ^3$ in the Minkowski spacetime
$\mathbb{R} ^{3,1}$, and $\Sigma _{l_0 +1}, \cdots, \Sigma _{l}$
can be isometrically embedded into $\mathbb{H} _{-\kappa _{l_0 +1}
^2} ^3, \cdots, \mathbb{H} _{-\kappa _{l} ^2} ^3$ in the Minkowski
spacetime $\mathbb{R} ^{3,1}$ respectively, we define the {\it
quasi local mass of $\Sigma $} as
 \beq
{\hat m}(\Sigma )=\sum _{1 \leq i \leq l _0} m(\Sigma _{i}) +\sum
_{l_0 +1 \leq i \leq l} {\hat m}(\Sigma _{i}).
\label{total-local-mass-new}
 \eeq

 \begin{theorem}\label{thm2}
Let $(N, {\widetilde g})$ be a spacetime which satisfies the
dominant energy condition. Let $(M, g, p)$ be a smooth initial
data set with the boundary $\Sigma$ which has finitely many
multi-components $\Sigma _i $, each of which is topological
2-sphere. Suppose that some $\Sigma _{i_0} $ has negative Gauss
curvature and let $K _{\Sigma _{i_0}} \geq -\kappa ^2$ ($\kappa >
0$) where $-\kappa ^2$ is the minimum of the Gauss curvature. If
either condition $(i)$ or condition $(ii)$ holds, then
 \begin{enumerate}
 \item ${\hat m}(\Sigma _{i _0}) \geq 0$;
 \item that ${\hat m}(\Sigma _{i _0})=0$ implies the energy-momentum
of spacetime satisfies
 \beQ
T_{00} = |f| |\phi | ^2, \;\;\;\;T _{0i} = f \langle \phi, e_0 \cdot
e _i \cdot \phi \rangle
 \eeQ
along $M$, where $f$ is a real function, $\phi $ is the unique
solution of (\ref{existence1}) or (\ref{existence2}) for some
$\breve{\phi} \in \breve{\Xi}$.
 \item Furthermore, if $p _{ij}=0$, then
${\hat m}(\Sigma _{i _0})=0$ implies that $M$ is flat with connected
boundary; if $p _{ij}=\pm \kappa g _{ij}$, then ${\hat m}(\Sigma _{i
_0})=0$ implies that $M$ has constant curvature $-\kappa ^2$.
 \end{enumerate}
 \end{theorem}
\pf By Lemma \ref{lemma}, statements $(1)$, $(2)$ and the first part
of statement $(3)$ can be proved by the same argument as the proof
of Theorem 1 in \cite{Z2}. For the proof of the second part of the
statement $(3)$, the vanishing quasi local mass implies
 \beQ
\conn _i \phi \pm \frac{\kappa}{2} e_0 \cdot e _i \cdot \phi =0.
 \eeQ
Since $\conn _i ( e _0 \cdot \phi ) =e _0 \cdot \conn _i \phi$, we
find the $M$ has constant Ricci curvature with the scalar curvature
$-6 \kappa ^2$. Therefore $M$ has constant curvature $-\kappa ^2$
because the dimension is 3. \qed

{\footnotesize {\it Acknowledgements.} The author is indebted to
J.X. Hong for some valuable conversations.}


\begin{thebibliography}{99}

\bibitem{BY1}
J.D. Brown and J.W. York, {\it Quasilocal energy in general
relativity}, Mathematical aspects of classical field theory
(Seattle, WA, 1991), 129-142, Contemp. Math., 132, Amer. Math. Soc.,
Providence, RI, 1992.
\bibitem{BY2}
J.D. Brown and J.W. York, {\it Quasilocal energy and conserved
charges derived from the gravitational action}, Phys. Rev. D(3) 47,
1407-1419 (1993).
\bibitem{CW}
M.P. do Carmo and F.W. Warner, {\it Rigidity and convexity of
hypersurfaces in spheres}, J. Diff. Geom. 4, 133-144 (1970).
\bibitem {CH} P. Chru\'sciel, M. Herzlich, {\it The mass of
asymptotically hyperbolic Riemannian manifolds}, Pacific J. Math.,
212, 231-264 (2003).
\bibitem{GL}
F. Guan and Y.Y. Li, {\it The Weyl problem with nonnegative Gauss
curvature}, J. Diff. Geom., 39, 331-342 (1994).
\bibitem{HZ}
J.X. Hong and C. Zuily, {\it Isometric embedding of the 2-sphere
with nonnegative curvature in $\mathbb{R} ^3 $}, Math. Z., 219,
323-334 (1995).
\bibitem{LY1}
C-C.M. Liu and S.T. Yau, {\it Positivity of quasilocal mass}, Phys.
Rev. Lett. 90, 231102 (2003).
\bibitem{LY2}
C-C.M. Liu and S.T. Yau, {\it Positivity of quasilocal mass II}, J.
Amer. Math. Soc. 19, 181 (2006).
\bibitem{P}
A.V. Pogorelov, {\it Some results on surface theory in the large},
Adv. Math. 1, 191-264 (1964).
\bibitem{ST1}
Y. Shi and L-F. Tam, {\it Positive mass theorem and the boundary
behavior of compact manifolds with nonnegative scalar curvature},
J.Diff.Geom. 62, 79 (2002).
\bibitem{ST2}
Y. Shi and L-F. Tam, {\it Boundary behaviors and scalar curvature of
compact manifolds}, math/0611253.
\bibitem{SY1}
R. Schoen, S.T. Yau, {\it On the proof of the positive mass
conjecture in general relativity}, Commun. Math. Phys. 65, 45-76
(1979).
\bibitem{SY2} R. Schoen, S.T. Yau, {\it Proof of the positive mass
theorem. II}, Commun. Math. Phys. 79, 231-260 (1981).
\bibitem{WY}
M.T. Wang and S.T. Yau, {\it A generalization of Liu-Yau's
quasi-local mass}, math.DG/0602321.
\bibitem{W}
X. Wang, {\it The mass of asymptotically hyperbolic manifolds}, J.
Diff. Geom., 57, 273-299 (2001).
\bibitem {Z1} X.Zhang, {\it A definition of total energy-momenta
and the positive mass theorem on asymptotically hyperbolic
3-manifolds I}, Commun. Math. Phys., 249, 529-548 (2004).
\bibitem{Z1-1}
X. Zhang, {\it The positive mass theorem near null infinity},
Proceedings of ICCM 2004, December 17-22, Hong Kong (eds. S.T.
Yau, etc.), AMS/International Press, Boston, to appear,
math/0604154.
\bibitem{Z2}
X. Zhang, {\it A new quasi-local mass and positivity}, Acta
Mathematica Sinica, English Series, to appear.
\end{thebibliography}
\end{document}